\documentstyle[12pt]{article}
\newcommand{\bm}[1]{\mbox{\boldmath $#1$}}
\newcommand{\be}{\begin{equation}}
\newcommand{\ee}{\end{equation}}
\newcommand{\rd}{{\rm d}}
\newcommand{\bb}[1]{\bibitem{#1}}
\begin{document}
\begin{titlepage}
\setcounter{page}{1}
\title{Particle-like solutions to topologically massive
 gravity}
\author{G\'erard Cl\'ement\thanks{E-mail:
 GECL@CCR.JUSSIEU.FR.} \\
\small Laboratoire de Gravitation et Cosmologie Relativistes
 \\
\small Universit\'e Pierre et Marie Curie, CNRS/URA769 \\
\small Tour 22-12, Bo\^{\i}te 142 \\
\small 4, place Jussieu, 75252 Paris cedex 05, France}
\bigskip
\date{\small March 25, 1994}
\maketitle
\begin{abstract}
The solution of topologically massive gravity with
 cosmological constant is reduced, for space-times with two
 commuting Killing vectors, to a special-relativistic
 dynamical problem. This approach is applied to the
 construction of a class of exact sourceless, horizonless
 solutions asymptotic to the BTZ extreme black holes.
\end{abstract}
\end{titlepage}
Three-dimensional Einstein gravity with a negative
 cosmological constant has recently been shown by Ba\~nados,
 Teitelboim and Zanelli (BTZ) \cite{1} to admit black-hole
 solutions which are in some respects quite similar to the
 Kerr black holes of four dimensional general relativity.
 There is however an important difference between the two
 theories: the field equations of three-dimensional
 cosmological gravity imply that the curvature is constant
 outside sources, so that the theory is dynamically trivial.
 To remedy this defect, Deser, Jackiw and Templeton \cite{2}
 have proposed a theory of three-dimensional gravity with a
Chern-Simons topological self-coupling --- topologically
 massive gravity (TMG) --- which admits massive excitations.
 The field equations of TMG with cosmological constant are
\be
G^{\mu}_{\;\;\nu}+\frac{1}{\mu} \, C^{\mu}_{\;\;\nu} =
 \Lambda \, \delta^{\mu}_{\;\;\nu}\;,
\ee
where $G^{\mu}_{\;\;\nu} \equiv R^{\mu}_{\;\;\nu}-
\frac{1}{2} \,R\,\delta^{\mu}_{\;\;\nu}$ is the Einstein
 tensor,
\be
C^{\mu}_{\;\;\nu} \equiv
 \varepsilon^{\mu\alpha\beta}\,D_{\alpha}\,(R_{\beta\nu}-
\frac{1}{4}\,g_{\beta\nu}\,R)
\ee
is the Cotton tensor (with $\varepsilon^{\mu\alpha\beta}$
 the antisymmetrical tensor), $\mu$ is the topological mass
 constant (we shall assume without loss of generality
 $\mu>0$) and $\Lambda$ is the cosmological constant. As
 remarked by Kaloper \cite{3}, these equations are trivially
 solved by the BTZ black-hole metric, which has a
 covariantly constant Ricci tensor, leading to a vanishing
 Cotton tensor\footnote{Note that this holds only because
 the BTZ black hole is everywhere regular. The well-known
 point-particle solution to three-dimensional gravity
 \cite{13} solves TMG only when the mass $m$ and spin
 $\sigma$ of the source are constrained by $m+\mu\sigma=0$
 \cite{4,5,6}.}.

Is there a non-trivial TMG analogue of the four-dimensional
 Kerr solution? Let us first make more precise what we mean
 by a Kerr analogue. We would like such a solution 1) to
 have an event horizon, and 2) for $\Lambda\leq0$, to be
 asymptotically particle-like, which we define as asymptotic
 to the conical metric of \cite{13} for $\Lambda=0$, or
 asymptotic to the BTZ metric for $\Lambda<0$.
 The black-hole solution of TMG with cosmological constant
 recently given by Nutku \cite{7} satisfies only the first
 of these criteria; this solution (eq.\  (18) below), which
 is a
generalization of the $\Lambda=0$ Vuorio solution \cite{9},
 does have an event horizon for a certain range of parameter
 values, but is not asymptotically particle-like. On the
 other hand, two kinds of asymptotically particle-like
 wormhole solutions have been found in TMG with vanishing
 cosmological constant \cite{8}, but these do not have
 horizons. In this Letter we shall construct a family of
 asymptotically particle-like solutions to TMG with negative
 cosmological constant. These new solutions turn out not to
 have a regular horizon, and so do not satisty our first
 criterion. However a subclass of these space-times are
 geodesically complete, and asymptotic to the BTZ extreme
 black holes, which we feel qualifies them as candidates for
 particle-like solutions of TMG. After all, what is the
 necessity of a horizon if there is no singularity to hide?

We first present a new, powerful approach to reduce the
 solution of TMG with two commuting Killing vectors $K_1$,
 $K_2$ to a dynamical problem in an abstract three-
dimensional Minkowski space. This approach is based on the
 observation \cite{8,12} that the $SL(2, R)$ group of
 transformations in the plane $(K_1, K_2)$ is locally
 isomorphic to the Lorentz group $SO(2, 1)$, which suggests
 the parametrization of the ``physical'' three-dimensional
 metric
\be
\rd s^2=\lambda_{ab}(\rho) \, \rd x^a \, \rd x^b + \zeta^{-
2} (\rho) \, R^{-2}(\rho) \, \rd \rho^2 ,
\ee
where $\lambda$ is the $2 \times 2$ matrix
\be
\lambda = \left( \begin{array}{cc}
   T+X & Y \\
    Y & T-X \end{array} \right),
\ee
$\det \lambda = R^2 \equiv \bm{X}^2$ is the Minkowski
 pseudo-norm of the vector $\bm{X}=(T,X,Y)$,
\be
\bm{X}^2 \equiv T^2-X^2-Y^2
\ee
(special linear transformations of $\lambda$ correspond to
 Lorentz rotations of $\bm{X}$), and the function
 $\zeta(\rho)$ allows for arbitrary reparametrizations of
 the coordinate $\rho$. If the vector $\bm{X}$ is space-like
 ($R^2 < 0$),
 the physical metric (3) has the Minkowskian signature, and
 is stationary rotationally symmetric if the orbits of the
 space-like Killing vector are closed.

The action for TMG with cosmological constant $\Lambda$ is
\be
I=I_E+I_{CS},
\ee
where
\be
I_{E} = -m \, \int \rd ^3 x \, \sqrt{|g|} \,\, (g^{\mu\nu}
 \, R_{\mu\nu}+2 \, \Lambda)
\ee
is the action for cosmological gravity ($\kappa = 1/2m$ is
 the Einstein gravitational constant), and
\be
I_{CS} = -\frac{m}{2\mu} \, \int \rd ^3 x \,
 \varepsilon^{\lambda\mu\nu} \, \Gamma^{\rho}_{\lambda\sigma}
 \, (\partial_\mu \Gamma^{\sigma}_{\rho\nu}+\frac{2}{3} \,
 \Gamma^{\sigma}_{\mu\tau} \, \Gamma^{\tau}_{\nu\rho})
\ee
is the Chern-Simons term. The action (6) reduces with the
 parametrization (3) to
\be
I = \int \rd ^2 x \int \rd  \rho \, L,
\ee
where the generalized Lagrangian \cite{14}
\be
L = - \frac{m}{2} \, \zeta \, \dot{\bm{X}}^2-2 \, m \,
 \zeta^{-1} \, \Lambda + \frac{1}{2\mu m} \, \zeta^2 \,
 (\bm{X},\dot{\bm{X}},\ddot{\bm{X}})
\ee
depends on the coordinates $X^i$, velocities $\dot{X}^i=\rd
 X^i/\rd \rho$, accelerations $\ddot{X}^i$ and on the
 dynamical variable $\zeta$, which acts as a Lagrange
 multiplier (in(10), $(\bm{X},\dot{\bm{X}},\ddot{\bm{X}})$
 is the mixed product $\varepsilon_{ijk}X^i \dot{X}^j
 \ddot{X}^k$).
 Varying this Lagrangian with respect to the
 coordinates $X^i$ leads to the equations of motion
\be
\ddot{\bm{X}} = - \frac{1}{2\mu} \, (3 \,\dot{\bm{X}}
 \bm{\wedge} \ddot{\bm{X}} + 2 \, \bm{X} \bm{\wedge}
 \dot{\ddot{\bm{X}}}).
\ee
These dynamical equations do not involve the cosmological
 constant, which occurs only in the constraint
\be
H \equiv -\frac{1}{2} \, m \, \dot{\bm{X}}^2 + \frac{m}{\mu}
 \, (\bm{X},\dot{\bm{X}},\ddot{\bm{X}}) = -2 \, m \, \Lambda
\ee
obtained by varying the Lagrangian (10) with respect to
 $\zeta$ (in (11) and (12), we have set the arbitrary
 function $\zeta$
 equal to 1 after variation). Equations (11) and (12) are
 equivalent to the field equations (1) in the case of a
 metric depending only on one variable.

The action (9) is by construction invariant under pseudo-
rotations and under translations in `time' $\rho$. The
 invariance under rotations leads to the conservation of
 generalized angular momentum
\be
\bm{J} = m \, \bm{X}\wedge\dot{\bm{X}} + \frac{m}{2\mu} \,
 [2 \, \bm{X}\wedge(\bm{X}\wedge\ddot{\bm{X}})-
\dot{\bm{X}}\wedge(\bm{X}\wedge\dot{\bm{X}})],
\ee
while the constant of the motion associated with
 translations of $\rho$ is the generalized hamiltonian $H$
 occurring in the left hand-side of eq.\  (12), which is of
 course the hamiltonian constraint for TMG.

Solutions of Einstein cosmological gravity ($\mu \rightarrow
 \infty$) depending on only one variable are geodesics of
 our abstract Minkowski space,
\be
\bm{X} = \bm{\alpha} \, \rho + \bm{\beta},
\ee
with $\bm{\alpha}^2 = 4 \Lambda$ from the Hamiltonian
 constraint; obviously these also solve the TMG equations
 (11), (12) for finite $\mu$. In this Letter we make the
 following ansatz for solutions to TMG:
\be
\bm{X} = \bm{\alpha} \, \rho + \bm{\beta} + \bm{\gamma} \,
 \rho^p
\ee
($\rho>0, p \,\, {\rm real}, p \neq 0, 1$). Then eqs.\  (11)
 imply that
\be
\bm{\alpha} \wedge \bm{\gamma} = -\frac{2\mu}{2p-1} \,
 \bm{\gamma}
\ee
($p \ne 1/2$), and either 1) $p=2$, or 2) $\bm{\beta} \wedge
 \bm{\gamma} = 0$. Taking into account the relations
 $\bm{\alpha}^2 = -4\mu^2/(2p-1)^2$, $\bm{\alpha} \cdot
 \bm{\gamma} = \bm{\gamma}^2 =0$ which follow from (16), the
 Hamiltonian constraint (12) then leads to the relation
\be
\frac{\mu^2}{(2p-1)^2} - \frac{p \, (p-1)}{2p-1} \,
 (\bm{\beta} \cdot \bm{\gamma}) \, \rho^{p-2} = -\Lambda.
\ee

In the first case ($p=2$), the solution depends a priori on
 five parameters, three for the vectors $\bm{\alpha}$,
 $\bm{\gamma}$ related by (16), two for the vector
 $\bm{\beta}$ restricted by (17). However, as in the case of
 Einstein cosmological gravity \cite{12}, stationary
 rotationally symmetric solutions really depend on only two
 parameters, as three of the parameters may be chosen at
 will by suitable metric reparametrizations involving time
 dilations, uniform frame rotations, and translations of
 $\rho$ (under which the ansatz (15) with $p=2$ is form-
invariant). The parametrization
\[
\bm{\alpha} = (-1,1,-\frac{2\mu}{3}), \, \, \, \bm{\beta} =
 (a,a,\frac{3b}{\mu}), \, \, \, \bm{\gamma} =
 \frac{\mu^2+9\Lambda}{12a} \, (1,-1,0)
\]
leads to the metric
\begin{eqnarray}
\rd s^2 = 2 \, a \, \rd t^2 &-& 2 \, \left( \frac{2\mu}{3}
 \, \rho - \frac{3b}{\mu} \right) \, \rd t \, \rd \theta
 \,\, + \,\, \left( \frac{\mu^2+9\Lambda}{6a} \, \rho^2 -
 2\rho \right) \, \rd \theta^2 \nonumber \\
 &-& \frac{\displaystyle \rd \rho^2}{\displaystyle
 {\frac{\displaystyle \mu^2-27\Lambda}{\displaystyle 9}} \,
 \rho^2 + 4 \, (a-b) \, \rho + \frac{\displaystyle
 9b^2}{\displaystyle \mu^2}} \, ,
\end{eqnarray}
which reduces to Nutku's solution \cite{7} if one defines
 $M=12(b-a)$, $J=6b$, $r^2=2\rho$ (and takes care that the
 cosmological constant in \cite{7} is $-\Lambda$). This
 generalized Vuorio solution has horizons for certain ranges
 of parameter values, but is obviously not asymptotically
 particle-like.

The second possibility ($\bm{\beta}\wedge\bm{\gamma}=0$)
 leads to $\bm{\beta} \cdot \bm{\gamma} = 0$ ($\bm{\gamma}$
 is null), so that the constraint (17) can be satisfied only
 if the cosmological constant is negative. Putting
 $\Lambda=-l^{-2}$ ($l>0$), this constraint selects the two
 possible values of the exponent $p$,
\be
p_{\pm} = \frac{1}{2} \, (1 \pm \mu \, l).
\ee
The constant angular momentum (13) is then related to the
 null vector $\bm{\beta}$ by
\be
\bm{J} = \frac{2m}{\mu l^2} \, (1 \pm \mu \, l) \,
 \bm{\beta}.
\ee
Now the general solution depends a priori on four arbitrary
 parameters (three for the vectors $\bm{\alpha}$,
 $\bm{\gamma}$ related by (16), one for the constant $c$
 determining $\bm{\beta} = c^{-1} \bm{\gamma}$), but the
 ansatz (15) is not form-invariant under translations of
 $\rho$, so that stationary rotationally symmetric solutions
 depend again on two parameters. We choose the
 parametrization
\[
\bm{\alpha} = l^{-2} \, (1-l^2,1+l^2,0), \,\,\, \bm{\beta} =
 -\frac{M}{4} \, (1+l^2,1-l^2,\mp 2l), \,\,\, \bm{\gamma} =
 c \bm{\beta},
\]
such that the metric
\begin{eqnarray}
\rd s^2 & = & \left( 2 \, l^{-2} \, \rho -
 \frac{m_{\pm}(\rho)}{2} \right) \, \rd t^2 \, \, \pm \, \,
 l \, m_{\pm}(\rho) \, \rd \theta \, \rd t \nonumber \\
& & - \mbox{} \left( 2 \, \rho + l^2 \,
 \frac{m_{\pm}(\rho)}{2} \right) \, \rd \theta^2 \, \, - \,
 \, \frac{l^2 \rd \rho^2}{4\rho^2}
\end{eqnarray}
with
\be
m_{\pm}(\rho) = M \, (1 + c \, \rho^{p_{\pm}})
\ee
reduces to the BTZ extreme black hole \cite{1} for $c=0$,
 $M>0$.

The solutions (21) are asymptotically particle-like if
 $p_{\pm}<0$, which occurs only for the exponent $p_-$ in
 the case $\mu l>1$. The corresponding solutions are
 asymptotic to the BTZ extreme black-hole metric for $M>0$,
 or to the corresponding extreme point-particle metric for
 $M<0$. The metric (21) is apparently singular at $\rho=0$,
 which is at
 infinite proper distance. To check whether $\rho=0$ is also
 at infinite geodesic distance, we write the first integral
 of the (physical) geodesic equation
\be
R^{-2} \, \left( \frac{\rd  \rho}{\rd  \tau} \right)^2 +
 \Pi^T \, \lambda^{-1} \, \Pi = \varepsilon,
\ee

where $\Pi$ is a constant column matrix, $\tau$ is an affine
 parameter, and $\varepsilon = +1, \, 0$ or $-1$. With the
 parametrization (5), eq.\  (23) simplifies to
\be
\left( \frac{\rd \rho}{\rd \tau} \right)^2 + \bm{P} \cdot
 \bm{X} - \varepsilon \, R^2 = 0,
\ee
where \bm{P} is a constant null vector. In the present case
 this equation reads
\be
\left( \frac{\rd \rho}{\rd \tau} \right)^2 + \bm{\alpha}
 \cdot \bm{P} \, \rho + \bm{\beta} \cdot \bm{P} \, (1+c \,
 \rho^{p_{\pm}}) +
 \varepsilon \, \frac{4\rho^2}{l^2} = 0.
\ee
For $p_- <0$, the effective potential near $\rho=0$ is
 dominated by the term $\bm{\beta} \cdot \bm{P} c \rho^p$,
 which is attractive or repulsive depending on the sign of
 $cM$. For $cM>0$, the constant $c \bm{\beta} \cdot \bm{P}$
 is generically negative, so that
 $\rho=0$ is at finite geodesic distance; unless $c=0$ (in
 which case $\rho=0$ is the horizon of the BTZ extreme black
 hole), the geodesics cannot for $p_-$ real be extended
 to negative $\rho$, so the space-time of metric (21) is
 singular{\samepage \footnote{This space-time is singular
 even though the curvature invariants $R=6/l^2, \,
 R^{\mu\nu}R_{\mu\nu}=12/l^4$ are regular (another instance
 in TMG is the exact $f=0$ solution of \cite{8}, eq.\  (27)
 below). Indeed,
 the singularity is due to the term $\bm{\gamma}\rho^p$ in
 (15), which does not contribute to the curvature invariants
 because the vector $\bm{\gamma}$ is null and orthogonal to
 $\bm{\alpha}$ and $\bm{\beta}$.}} on the circle $\rho=0$.
 On the other hand, for $cM<0$, $c \bm{\beta} \cdot \bm{P}
 \geq
 0$ and $\rho$ can attain the value $0$ only if the null
 vectors $\bm{P}$ and $\bm{\beta}$ are collinear (implying
 also $\bm{\alpha} \cdot \bm{P} = 0$) and $\varepsilon = -
1$; the time-like circle $\rho = 0$ is at infinite affine
distance on these space-like geodesics, so that the geometry
(21) is regular (as in the case of the extreme point-
particle solution of Einstein cosmological gravity
 \cite{12}).

For completeness, we also briefly discuss the case
 $p_{\pm}>0$, which occurs for $p_+$ if $\mu l>1$, or for
 both exponents $p_{\pm}$ if $\mu l<1$. Now the effective
 potential near $\rho =0$ is dominated by the constant
 $\bm{\beta} \cdot \bm{P}$. We do not repeat the discussion,
 which follows
 closely that made in the case $p_- <0$, leading to the
 conclusion that the geometry (21) is singular for $M>0$,
 and regular for $M<0$.

The solution (21) is such that the vector $\bm{X} (\rho)$
 evolves in the plane $(\bm{\alpha},\bm{\beta})$ orthogonal
 to the constant vector $\bm{J}$. One may inquire whether
 there are more general planar solutions
\be
\bm{X} = \bm{\alpha} \, F(\rho) + \bm{\beta} \, G(\rho)
\ee
to cosmological TMG. We have found that this planar family
 of solutions includes, besides the trivial solution
 (14) ($\Lambda$ real) and the solution (21) ($\Lambda<0$),
 only one other solution, the exact $f=0$ solution of
 \cite{8} ($\Lambda=0$),
\be
\bm{X} = \bm{\alpha} + \bm{\beta} \, (c \, \rho + e^{-\nu \rho}),
\ee
which has properties somewhat similar to those of solution
 (21), and is particle-like for $c=0$.

We have reduced the solution of TMG with two commuting
 Killing vectors to a dynamical problem in an abstract
 three-dimensional Minkowski space. A polynomial ansatz
 leads to two classes of solutions. The Vuorio-Nutku
 solutions of the first class ($p=2$) are not aymptotically
 particle-like. The second class ($p = \frac{1}{2} (1 \pm
 \mu l)$) includes a subclass of regular solutions
 asymptotic to the BTZ extreme black holes --- sourceless,
 horizonless particle-like solutions of TMG with negative
 cosmological constant.

\vspace{5 mm}
\large \noindent Acknowledgement
\vspace{5 mm}

\normalsize \noindent I thank S. Deser for stimulating
 discussions and helpful suggestions.

\end{document}